\documentclass[11pt,fleqn,twoside]{article}

\usepackage{amsthm}
\makeatletter
\newcommand{\prava}[1]{\small\it
\begin{flushleft}
Copyright \copyright \ 2000 by  #1
\end{flushleft}}

\newcommand{\name}[1]{\begin{flushleft}
                       \LARGE \bf #1
                       \end{flushleft}\vspace{-3mm}}

\newcommand{\Author}[1]{\begin{flushleft}
                       \it #1 \end{flushleft}}

\newcommand{\Adress}[1]{\begin{flushleft}
                       \it #1 \end{flushleft}}

\newcommand{\Date}[1]{\begin{flushleft}
                      \small  \it #1 \end{flushleft}}

\newcommand{\ehkol}{Author \ name}
\newcommand{\ohkol}{Article \ name}

\renewcommand{\@evenhead}{
\hspace*{-3pt}\raisebox{-15pt}[\headheight][0pt]{\vbox{\hbox to \textwidth 
{\thepage \hfil \ehkol}\vskip4pt \hrule}}}
\renewcommand{\@oddhead}{
\hspace*{-3pt}\raisebox{-15pt}[\headheight][0pt]{\vbox{\hbox to \textwidth 
{\ohkol \hfil \thepage}\vskip4pt\hrule}}}
\renewcommand{\@evenfoot}{}
\renewcommand{\@oddfoot}{}

\long\def\@makecaption#1#2{%
  \vskip\abovecaptionskip
  \sbox\@tempboxa{\small \textbf{#1.}\ \ #2}%
  \ifdim \wd\@tempboxa >\hsize
    {\small \textbf{#1.}\ \ #2}\par
  \else
    \global \@minipagefalse
    \hb@xt@\hsize{\hfil\box\@tempboxa\hfil}%
  \fi
  \vskip\belowcaptionskip}

\def\numberwithin#1#2{\@ifundefined{c@#1}{\@nocounterr{#1}}{%
  \@ifundefined{c@#2}{\@nocnterr{#2}}{%
  \@addtoreset{#1}{#2}%
  \toks@\@xp\@xp\@xp{\csname the#1\endcsname}%
  \@xp\xdef\csname the#1\endcsname
    {\@xp\@nx\csname the#2\endcsname
     .\the\toks@}}}}

\newcommand{\resetfootnoterule} {
  \renewcommand\footnoterule{%
  \kern-3\p@
  \hrule\@width.4\columnwidth
  \kern2.6\p@}
}

\makeatother

\numberwithin{equation}{section}

\def\+{\;+\;}
\def\-{\;-\;}
\def\*{\,\cdot\,}
\def\p{\mbox{\sf p}}

\def\u{\mbox{\sf u}}
\def\w{\mbox{\sf w}}

\def\o{\mbox{\bf $\omega$}}
\def\ds{\displaystyle}
\def\R{\mbox{\bf R}}
\def\U{\mbox{\bf U}}
\newtheorem{theorem}{Theorem}
\newtheorem{prop}{Proposition}

\begin{document}

\thispagestyle{empty}
\renewcommand{\ehkol}{S.M.\ Sergeev}
\renewcommand{\ohkol}{On Exact Solution of a Classical 3D Integrable Model}

\begin{flushleft}
\footnotesize \sf
Journal of Nonlinear Mathematical Physics \qquad 2000, V.7, N~1,
\pageref{sergeev_fp}--\pageref{sergeev_lp}.
\hfill {\sc Article}
\end{flushleft}

\vspace{-5mm}

\renewcommand{\footnoterule}{}
{\renewcommand{\thefootnote}{}
 \footnotetext{\prava{S.M.\ Sergeev}}}

\name{On Exact Solution of a\\ Classical 3D
Integrable Model} \label{sergeev_fp}

\Author{S.M.\ SERGEEV}

\Adress{Branch Institute for Nuclear Physics, 
Protvino 142284, Russia \\ 
E-mail: sergeev\_ms@mx.ihep.su}

\Date{Received July 23, 1999; Revised October 2, 1999; Accepted
November 3, 1999}

\begin{abstract}
\noindent
We investigate some classical evolution model
in the discrete $2+1$ space-time. A map, giving an one-step
time 
evolution, may be derived as the compatibility condition
for some systems of linear equations for a set of auxiliary
linear variables. Dynamical variables for the evolution model
are the coefficients of these systems of linear equations.
Determinant of any system of linear equations is a polynomial
of two numerical quasimomenta of the auxiliary linear variables.
For one, this determinant is 
the generating functions of all integrals
of motion for the evolution, 
and on the other hand it 
defines a high genus algebraic curve.
The dependence of the dynamical variables on the space-time point
(exact solution) may be 
expressed in terms of theta functions on the jacobian 
of this curve. This is the main result of our paper.
\end{abstract}

\section{Introduction}

In this paper we give an exact solution of a classical
evolution model in  discrete $2+1$ space-time.
This model was formulated in \cite{s-3dsympl}.
The map of dynamical variables, 
governing the one-step evolution, was derived
as the compatibility condition for two sets of linear relations,
associated with the usual graphical representation
of left and right hand sides of the famous Yang -- Baxter 
equation. Such form of the zero curvature condition
generalizes the usual concept of the local Yang -- Baxter
equation as the zero curvature condition for 3d lattice models. 
Linear variables may be assigned
either to the vertices or to the sites of the 
two dimensional auxiliary graphs (these
two assignments are the dual ones), and the 
coefficients of the linear relations -- nothing but the 
dynamical variables -- are to be assigned to the vertices.
Main feature of the map of the dynamical variables
is that it is canonical
with respect to local Poisson brackets and thus may be easily
quantized \cite{s-qem}, so that the
auxiliary linear systems exist even in the quantum case
\cite{electric}. 

Evolution models arise when one considers flat regular
graphs, formed by straight lines, on a two dimensional torus.
The only demand is that the graph must contain Yang-Baxter
triangles, so that a simultaneous ``bypass of some lines through
appropriate vertices'' will restore the geometry of the graph.
Thus, with the map associated with the Yang -- Baxter type 
triangles, there appears the map of the dynamical variables,
assigned to the vertices of the primary lattice, into
the set of the dynamical variables, assigned in fact to the
same lattice. This gives the one-step evolution of the
discrete $2+1$ evolution model. 

Simple square lattice does not fit the demands,
because of it does not contain the triangles.
The simplest two dimensional
graph with the properties demanded is the so called kagome lattice
\footnote{``Kagome'' is not a name, it is a kind of 
Japanese mats.}
(see the figures below). This is not something strange,
the 2d kagome lattice is nothing but a section of a regular
3d cubic lattice by an inclined plane.

The governing map was derived for the open linear system
(i.e. the number of the equations is less then the number
of auxiliary linear variables), but nevertheless
the linear systems may be written for the 2d lattice
with the toroidal boundary conditions. Also, in general,
dealing with the linear equations, one may impose on the
linear variables quasiperiodical boundary conditions
in both directions of the torus with two numerical
quasimomenta. In this case
the number of the auxiliary linear variables
coincides with the number of the linear equations
and one may ask for the admissibility of the linear system,
i.e. for the zero value of corresponding determinant.
Because of the evolution arises as a simple compatibility
of two similar linear systems, the admissibility
of the primary system provides the admissibility
of the evaluated one. Therefore the determinant
is at least an ideal of the evolution. Moreover,
being normalized appropriately, formal determinant $J(A,B)$
of the linear system as a polynomial of the quasimomenta $A$ and $B$
is conserved by one-step evolution map, i.e.
is the generating function of the integrals of motion.
All these remains valid and in the quantum case, where
$J(A,B)$ is an operator-valued functional.

In our classical case and finite spatial size
of the two dimensional lattice, equation $J(A,B)=0$ gives a finite
genus $g$ algebraic curve $\Gamma$, so that the integrals
of motion are interpreted as moduli of $\Gamma$.
With $\Gamma$ and a bit of additional
information concerning the initial state given,
the system of the auxiliary
linear variables may be easily parametrized as the
meromorphic functions on $\Gamma$ in terms of the theta
functions on $\mbox{Jac}\;\Gamma$. 
Doing this, we get at once 
the parametrization of the dynamical variables in terms
of theta functions on $\mbox{Jac}\;\Gamma$ and obtain the exact
solution.


\resetfootnoterule%
\noindent%
Perhaps it would be expedient to discuss several $3d$
discrete integrable models from the point of view of their linear
systems and indicate the place of the model being considered
among them. Spatial nature of any $3d$ integrable model
means that geometrically linear variables are assigned to
several elements of $3d$ cubic lattice. This assignment
gives a type of the linear system. 
Consider three main scenarios corresponding to three
relative integrable models.

The first, most simple type
\cite{Hirota,Miwa-Hirota,HM-croud,klwz-elliptic}: let the
vertices of the cubic lattice $Z^3$ have the coordinates
$\p=(a,b,c)$, $a,b,c\in Z$. Consider the even sublattice of
it, $Z^3_{even}$: $\p=(a,b,c)$, $a+b+c=\mbox{even}$.
Points $Z^3_{even}$ triangulize the three dimensional Euclidean 
space into the following convex bodies: set of
octahedra and sets of two types of tetrahedra (up to regular
translations). 
Assign the auxiliary linear variables to the
vertices of $Z^3_{even}$. Linear equations are assigned to
the triangles of the triangulation described. Primary set of
the linear equations maybe chosen for the triangles -- sides
of one of the tetrahedra. System of coefficients of the
linear equations gives the tau function for Hirota's
discrete bilinear equation (the ``{\em octahedron
equation}'' from this naive geometrical point of view).

The second type: assign the auxiliary linear variables to
the facets of $Z^3$. Linear relation are to be written
for each edge surrounded by four facets. This corresponds to
Korepanov's block-matrix models \cite{korepanov-diss}.
Hirota and Hirota-Miwa' equations are the simplest
compatibility conditions for Korepanov's linear problem
\cite{kks-fte,rmk-lybe,oneparam}.

The third type: dealing with the cubic lattice, most obvious
scenario is to assign the linear variables to all the
vertices of it (or, correspondingly, to the sites of the
dual lattice). Each linear equation corresponds to a square
(any face of the elementary cube). This is our case.
Independent Lagrangian -- type variables are a triplet of
``tau -- functions'', obeying the system of the ``{\em cube
equations}''. 
Taking a section of the cubic lattice by an inclined
plane, we at first get the kagome lattice geometrically,
and the linear system for it as a part of whole linear
relations secondly.
Details may be found in
\cite{s-3dsympl,s-qem,s-tau}. The advantages of this
approach (Poisson structure, quantization etc.) were
mentioned at the beginning of this introduction.

Few remarks concerning the section by the inclined plane
and the evolution. For the cubic lattice
$\p=(a,b,c)$, $a,b,c\in Z$, the sections mentioned 
are the planes $a+b+c=t=\mbox{const}$.
Equations of motion normally may be solved so that all the
dynamical variables for $a+b+c=t+1$ are expressed from the
dynamical variables for $a+b+c=t$. This gives the natural
notion of the evolution as the map from $t$ to $t+1$.
As it was mentioned, geometrically the section $a+b+c=t$ 
is the kagome lattice.

This paper is organized as follows. First, we recall the
formulation of the model, describing the dynamical system
and defining the evolution. Second, introducing the linear
system, we define the generating function for the integrals
of motion. All these are based on the results of
\cite{s-3dsympl,s-qem}, and we rather enumerate the facts.
In the fourth section we analyse the curve, parametrize the
auxiliary linear variables and derive the expressions for 
the dynamical variables. In the final section we
discuss possible applications of the results obtained.

\begin{figure}[tbp]
\centering
\setlength{\unitlength}{0.25mm} 
\thicklines
\begin{picture}(450,280)
\put(85,0){
\begin{picture}(280,280)
\multiput(20,0)(120,0){3}{\vector(0,1){280}}
\multiput(0,20)(0,120){3}{\vector(1,0){280}}
\put(0,100){\vector(1,-1){100}}
\put(0,220){\vector(1,-1){220}}
\put(60,280){\vector(1,-1){220}}
\put(180,280){\vector(1,-1){100}}
\put(147,147){\scriptsize $1$}
\put(82,147){\scriptsize $3$}
\put(147,82){\scriptsize $2$}
\put(110,117){\scriptsize $(a,b)$}
\put(95,247){\scriptsize $(a\!+\!1,b)$}
\put(215,127){\scriptsize $(a,b\!+\!1)$}
\put(283,232){\vector(-4,1){40}}
\put(288,227){\scriptsize $(a\!+\!1,b\!+\!1)$}
\end{picture}}
\end{picture}
\caption{Labelling of the kagome lattice.}
\label{fig-kagome}
\end{figure}
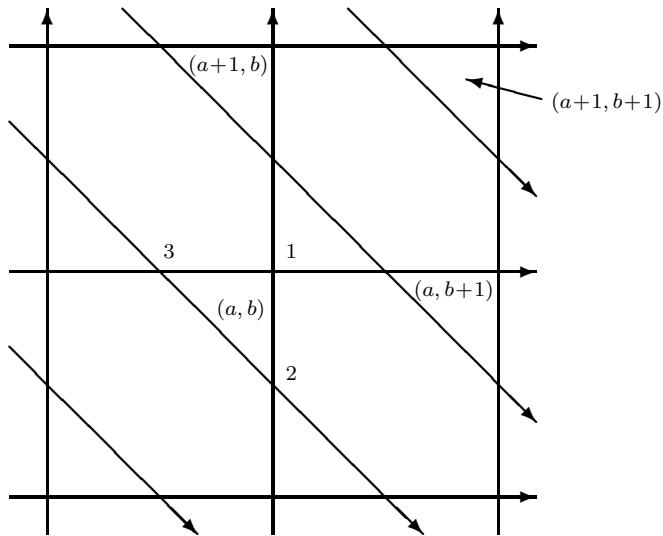

\section{Discrete evolution}

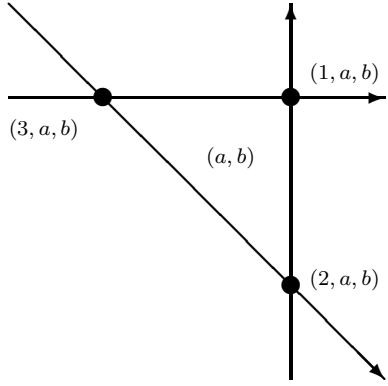
\begin{figure}[tbp]
\centering
\setlength{\unitlength}{0.25mm} 
\thicklines
\begin{picture}(450,200)
\put(125,0){
\begin{picture}(200,200)
\put( 150 ,   0 ){\vector( 0,1){200}}
\put(   0 , 150 ){\vector( 1,0){200}}
\put(   0 , 200 ){\vector(1,-1){200}}
\put( 150 , 150 ){\circle*{10}}
\put(  50 , 150 ){\circle*{10}}
\put( 150 ,  50 ){\circle*{10}}
\put(160,160){\scriptsize $(1,a,b)$}
\put(160,50){\scriptsize $(2,a,b)$}
\put(0,130){\scriptsize $(3,a,b)$}
\put(105,115){\scriptsize $(a,b)$}
\end{picture}}
\end{picture}
\caption{$(a,b)$ -- triangle.}
\label{fig-triangle}
\end{figure}

System of the dynamical variables assigned to a finite
$M\times M$ kagome lattice (see Figure~\ref{fig-kagome}) 
is a set of $3\;M^2$ pairs
\begin{equation}
\ds [\;\u_{j,a,b}\;,\;\w_{j,a,b}\;]\;,\;\;\;
j\;=\;1,2,3,\;\;a,b\;\in\;Z_M\;.
\end{equation}
Arrangement of the triangles of the kagome
lattice is shown in Figure~\ref{fig-kagome}.
Consideration of the lattice on a torus implies the
periodical boundary conditions for the dynamical variables
\begin{equation}
\ds
\u_{j,a+M,b}\;=\;\u_{j,a,b+M}\;=\;\u_{j,a,b}\;,\;\;\;
\w_{j,a+M,b}\;=\;\w_{j,a,b+M}\;=\;\w_{j,a,b}\;.
\end{equation}
Geometrically the indices $(j,a,b)$ are assigned to the
vertices of the kagome lattice, so that for $a,b$ given
$(j,a,b)$, $j=1,2,3$ mark three vertices of a definite
triangle, as it is shown in Figure~\ref{fig-triangle}. Whole
kagome lattice may be obtained as the up-down and left-right
translations of the triangle $(a,b)$. It is supposed that
$a$ increases to up and $b$ increases to right.

Impose the following Poisson structure on the set of the
dynamical variables:
\begin{equation}
\ds \left\{\;\u_{j,a,b}\;,\;\w_{j,a,b}\;\right\}\;=\;
\u_{j,a,b}\,\w_{j,a,b}\;,
\end{equation}
and any other Poisson bracket is zero. Remarkable is the
locality of the dynamical variables.

Evolution of the system is governed by the fundamental map
$\R$. Consider one isolated triangle $(a,b)$. Define map
\begin{equation}
\ds\R\;:\;[\u_{j},\w_{j}]\;\mapsto\;
[\u_j^\prime,\w_j^\prime]\;\;\;
j\;=\;1,2,3\,,\;\;\;(a,b)\;\;
\mbox{implied to be fixed,}
\end{equation}
by the following relations:
\begin{equation}\label{R-map}
\ds\begin{array}{@{}ll}
(i)&\ds\left\{\begin{array}{ccl}
\ds\u_1^\prime & = & \ds
{\kappa_2^{}\;\u_1^{}\;\u_2^{}\;\w_2^{}\over
\kappa_1^{}\;\u_1^{}\;\w_2^{}\;+\;\kappa_3^{}\;\u_2^{}\;\w_3^{}
\;+\;\kappa_1^{}\;\kappa_3^{}\;\u_1^{}\;\w_3^{}}\;,\\
&&\\
\ds\w_1^\prime & = & \ds
{\w_1^{}\;\w_2^{}\;+\;\u_3^{}\;\w_2^{}\;+\;
\kappa_3^{}\;\u_3^{}\;\w_3^{}\over\w_3^{}}\;,
\end{array}\right.\\
&\\&\\
(ii)&\ds\left\{\begin{array}{ccl}
\ds\u_2^\prime & = & \ds
{\u_1^{}\;\u_2^{}\;\u_3^{}\over\u_2^{}\;\u_3^{}\;+\;
\u_2^{}\;\w_1^{}\;+\;\kappa_1^{}\;\u_1^{}\;\w_1^{}}\;,\\
&&\\
\ds\w_2^\prime & = & \ds
{\w_1^{}\;\w_2^{}\;\w_3^{}\over\w_1^{}\;\w_2^{}\;+\;
\u_3^{}\;\w_2^{}\;+\;\kappa_3^{}\;\u_3^{}\;\w_3^{}}\;,
\end{array}\right.\\
&\\&\\
(iii)&\ds\left\{\begin{array}{ccl}
\ds\u_3^\prime & = & \ds {\u_2^{}\;\u_3^{}\;+\;
\u_2^{}\;\w_1^{}\;+\;
\kappa_1^{}\;\u_1^{}\;\w_1^{}\over\u_1^{}}\;,\\
&&\\
\ds\w_3^\prime & = & \ds 
{\kappa_2^{}\;\u_2^{}\;\w_2^{}\;\w_3^{}\over
\kappa_1^{}\;\u_1^{}\;\w_2^{}\;+\;
\kappa_3^{}\;\u_2^{}\;\w_3^{}\;+\;
\kappa_1^{}\;\kappa_3^{}\;\u_1^{}\;\w_3^{}}\;.
\end{array}\right.\end{array}\end{equation}
$\kappa_{1,2,3}^{}$ are arbitrary numbers (not the dynamical
variables). 

\begin{prop}
The map $\R$, (\ref{R-map}),
conserves the local Poisson structure,
\begin{equation}\label{localPoisson}
\ds
\{\u_{j}^{},\w_{j'}^{}\}\;=\;
\delta_{j,j'}^{}\;\u_{j}^{}\,\w_{j}^{}\;\;
\Leftrightarrow\;\;
\{\u_{j}^\prime,\w_{j'}^\prime\}\;=\;
\delta_{j,j'}^{}\;
\u_{j}^\prime\,\w_{j}^\prime\;,
\end{equation}
i.e. $\R$ is the canonical map.
\end{prop} 

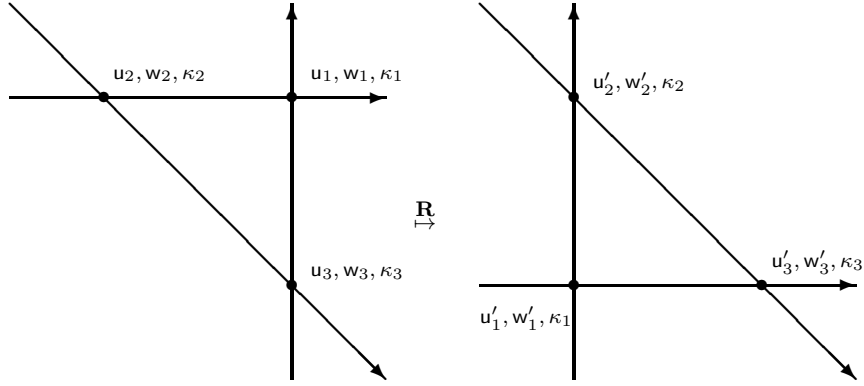
\begin{figure}[tb]
\centering

\setlength{\unitlength}{0.25mm} 
\thicklines
\begin{picture}(450,200)
\put(00,0){
\begin{picture}(200,200)
\put(0,150){\vector(1,0){200}}\put(50,150){\circle*{5}}
\put(0,200){\vector(1,-1){200}}\put(150,50){\circle*{5}}
\put(150,0){\vector(0,1){200}}\put(150,150){\circle*{5}}
\put(160,160){\scriptsize $\u_1,\w_1,\kappa_1$}
\put(55,160){\scriptsize $\u_2,\w_2,\kappa_2$}
\put(160,55){\scriptsize $\u_3,\w_3,\kappa_3$}
\end{picture}}
\put(220,80){\scriptsize $\ds\stackrel{\R}{\mapsto}$}
\put(250,0){
\begin{picture}(200,200)
\put(50,0){\vector(0,1){200}}\put(50,50){\circle*{5}}
\put(0,50){\vector(1,0){200}}\put(50,150){\circle*{5}}
\put(0,200){\vector(1,-1){200}}\put(150,50){\circle*{5}}
\put(0,30){\scriptsize $\u_1',\w_1',\kappa_1$}
\put(60,155){\scriptsize $\u_2',\w_2',\kappa_2$}
\put(155,60){\scriptsize $\u_3',\w_3',\kappa_3$}
\end{picture}}
\end{picture}
\caption{Geometrical representation of the map $\R$.}
\label{fig-map}
\end{figure}

Geometrically $\R$ may be
interpreted as the map from one Yang -- Baxter triangle
to another, as it is shown in Figure~\ref{fig-map},
i.e. as a bypass of a line through the opposite vertex.

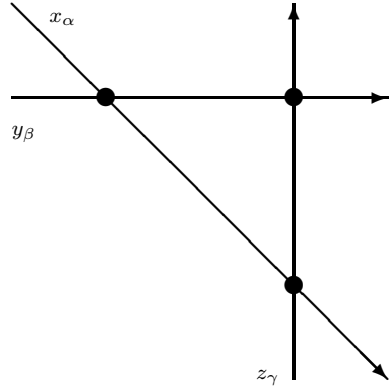
\begin{figure}[tb]
\centering
\setlength{\unitlength}{0.25mm} 
\thicklines
\begin{picture}(450,200)
\put(125,0){
\begin{picture}(200,200)
\put( 150 ,   0 ){\vector( 0,1){200}}
\put(   0 , 150 ){\vector( 1,0){200}}
\put(   0 , 200 ){\vector(1,-1){200}}
\put( 150 , 150 ){\circle*{10}}
\put(  50 , 150 ){\circle*{10}}
\put( 150 ,  50 ){\circle*{10}}
\put(20,190){\scriptsize $x_{\alpha}$}
\put(0,130){\scriptsize $y_{\beta}$}
\put(130,0){\scriptsize $z_{\gamma}$}
\end{picture}}
\end{picture}
\caption{Accessories of the lattice: lines.}
\label{fig-lines}
\end{figure}

Turn now to the whole lattice.
We will distinguish three types of lines with
respect to their slopes by the letters $x,y,z$
as it is shown in the Figure~\ref{fig-lines} and 
enumerate the lines,
$x_\alpha,y_\beta,z_\gamma$, $\alpha,\beta,\gamma\in Z_M$. 
``Spectral parameters''
$\kappa_{j,a,b}$ actually depend on numbers of the
corresponding lines.

Evolution of the lattice is the simultaneous shift of all
lines of one type in one direction. The result of such shift
is the application of $\R$ to each $(a,b)$ triangle and some
re-numeration of the images of
$\u_{j,a,b}^\prime,\w_{j,a,b}^\prime$. This re-numeration
depends on what type of vertices we leave immovable. Choose
the vertices of type $(1,a,b)$ motionless, i.e. we shift all
$x$ lines to the north-east direction. Define the functional
map $\U$, acting on the space of functions of the dynamical
variables $[\u_{j,a,b},\w_{j,a,b}]$:
\begin{equation}
\ds
\left(\U\,\circ\,f\right)(\;\u_{j,a,b}\;,\;\w_{j,a,b}\;)\;=\;
f(\;\U^*\,\circ\,\u_{j,a,b}\;,\;\U^*\,\circ\,\w_{j,a,b}\;)\;,
\end{equation}
where
\begin{equation}\label{ustar}
\ds\begin{array}{@{}ll}
\ds \U^*\,\circ\,\u^{}_{1,a,b}\;=\;\u^\prime_{1,a,b}\;,&
\ds \U^*\,\circ\,\w^{}_{1,a,b}\;=\;\w^\prime_{1,a,b}\;,\\
&\\
\ds \U^*\,\circ\,\u^{}_{2,a,b}\;=\;\u^\prime_{2,a+1,b}\;,&
\ds \U^*\,\circ\,\w^{}_{2,a,b}\;=\;\w^\prime_{2,a+1,b}\;,\\
&\\
\ds \U^*\,\circ\,\u^{}_{3,a,b}\;=\;\u^\prime_{3,a,b+1}\;,&
\ds \U^*\,\circ\,\w^{}_{3,a,b}\;=\;\w^\prime_{3,a,b+1}\;,
\end{array}\end{equation}
where, for example, $\u^\prime_{1,a,b}$ means that
we take the expression for $\u_1^\prime$ from (\ref{R-map})
and add the indices $a,b$ to each $\u_j$, $\w_j$ there.
Indices $(a+1,b)$ and $(a,b+1)$ in the second and third
lines of (\ref{ustar}) are the re-enumeration mentioned above.

Obviously, $\U$ conserves the Poisson brackets, and thus it
is the canonical map. $\U$ is identified with the one step
discrete evolution, so that if
\begin{equation}
\ds f(\;\u_{j,a,b}\;,\;\w_{j,a,b}\;)\;=\;f(t_0)\;,
\end{equation}
then
\begin{equation}
\ds f(t_0\;,\;t)\;=\;\left(\U^{t}\circ f\right)(t_0)\;.
\end{equation}

\section{Linear system and the integrals of motion}

Map $\R$ (\ref{R-map}) was ``derived'' in \cite{electric} as
a zero curvature condition for a system of linear equations.
There are at least two ways to define the linear system, 
and here we use the co-current form
according to the terminology of \cite{s-3dsympl,s-qem}. 

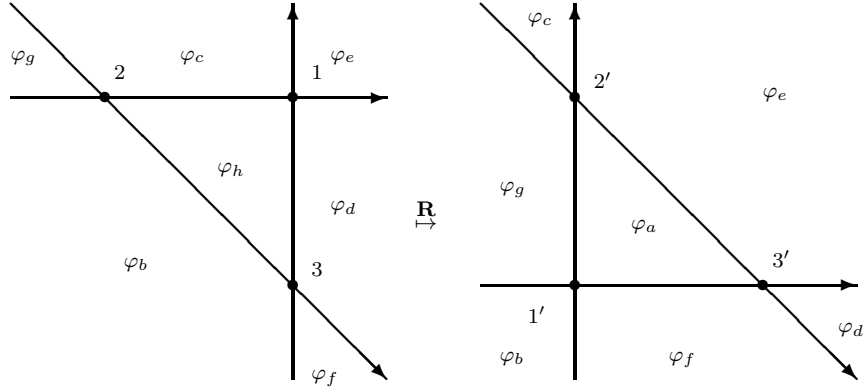
\begin{figure}[tb]
\centering

\setlength{\unitlength}{0.25mm} 
\thicklines
\begin{picture}(450,200)
\put(00,0){
\begin{picture}(200,200)
\put(0,150){\vector(1,0){200}}\put(50,150){\circle*{5}}
\put(0,200){\vector(1,-1){200}}\put(150,50){\circle*{5}}
\put(150,0){\vector(0,1){200}}\put(150,150){\circle*{5}}
\put(160,160){\scriptsize $1$}
\put(55,160){\scriptsize $2$}
\put(160,55){\scriptsize $3$}
\put(60,60){\scriptsize $\varphi_b$}
\put(0,170){\scriptsize $\varphi_g$}
\put(160,0){\scriptsize $\varphi_f$}
\put(90,170){\scriptsize $\varphi_c$}
\put(170,90){\scriptsize $\varphi_d$}
\put(170,170){\scriptsize $\varphi_e$}
\put(110,110){\scriptsize $\varphi_h$}
\end{picture}}
\put(220,80){\scriptsize $\ds\stackrel{\R}{\mapsto}$}
\put(250,0){
\begin{picture}(200,200)
\put(50,0){\vector(0,1){200}}\put(50,50){\circle*{5}}
\put(0,50){\vector(1,0){200}}\put(50,150){\circle*{5}}
\put(0,200){\vector(1,-1){200}}\put(150,50){\circle*{5}}
\put(25,30){\scriptsize $1'$}
\put(60,155){\scriptsize $2'$}
\put(155,60){\scriptsize $3'$}
\put(10,10){\scriptsize $\varphi_b$}
\put(10,100){\scriptsize $\varphi_g$}
\put(100,10){\scriptsize $\varphi_f$}
\put(25,190){\scriptsize $\varphi_c$}
\put(190,25){\scriptsize $\varphi_d$}
\put(150,150){\scriptsize $\varphi_e$}
\put(80,80){\scriptsize $\varphi_a$}
\end{picture}}
\end{picture}
\caption{Linear variables of the map $\R$.}
\label{fig-linears}
\end{figure}

We start from the linear system for isolated 
Yang -- Baxter triangles and explain how the map $\R$ appears. 
First, in addition to the vertex variables $\u_j,\w_j,\kappa_j$
introduce eight auxiliary variables $\varphi_a...\varphi_h$
living in the sites of the 2d graphs as it is shown in 
Figure~\ref{fig-linears}.

Please note that the variables $\varphi_b...\varphi_g$
are the same in the left and right hand sides of 
Figure~\ref{fig-linears} and belong to the equivalent open cells.
Left and right hand side graphs differ by $\varphi_h$
and $\varphi_a$. Consider now
two sets of linear relations: for the left hand side graph
\begin{equation}\label{leftset}
\ds\begin{array}{@{}ccl}
\ds 0\;=\;f_1^{} & \stackrel{def}{=} & \ds
\varphi_c^{}\;-\;\varphi_e^{}\,\u_1^{}\;+\;\varphi_h^{}\,\w_1^{}\;+\;
\varphi_d^{}\,\kappa_1^{}\,\u_1^{}\,\w_1^{}\;,\\
&&\\
\ds 0\;=\;f_2^{} & \stackrel{def}{=} & \ds
\varphi_h^{}\;-\;\varphi_d^{}\,\u_2^{}\;+\;\varphi_b^{}\,\w_2^{}\;+\;
\varphi_f^{}\,\kappa_2^{}\,\u_2^{}\,\w_2^{}\;,\\
&&\\
\ds 0\;=\;f_3^{} & \stackrel{def}{=} & \ds
\varphi_g^{}\;-\;\varphi_c^{}\,\u_3^{}\;+\;\varphi_b^{}\,\w_3^{}\;+\;
\varphi_h^{}\,\kappa_3^{}\,\u_3^{}\,\w_3^{}\;,
\end{array}
\end{equation}
and for the right hand side graph
\begin{equation}
\ds\begin{array}{@{}ccl}\label{rightset}
\ds 0\;=\; f_1^{\prime} & \stackrel{def}{=} & \ds
\varphi_g^{}\;-\;\varphi_a^{}\,\u_1^{\prime}\;+\;
\varphi_b^{}\,\w_1^{\prime}\;+\;
\varphi_f^{}\,\kappa_1^{}\,\u_1^{\prime}\,\w_1^{\prime}\;,\\
&&\\
\ds 0\;=\;f_2^{\prime} & \stackrel{def}{=} & \ds
\varphi_c^{}\;-\;\varphi_e^{}\,\u_2^{\prime}\;+\;
\varphi_g^{}\,\w_2^{\prime}\;+\;
\varphi_a^{}\,\kappa_2^{}\,\u_2^{\prime}\,\w_2^{\prime}\;,\\
&&\\
\ds 0\;=\;f_3^{\prime} & \stackrel{def}{=} & \ds
\varphi_a^{}\;-\;\varphi_e^{}\,\u_3^{\prime}\;+\;
\varphi_f^{}\,\w_3^{\prime}\;+\;
\varphi_d^{}\,\kappa_3^{}\,\u_3^{\prime}\,\w_3^{\prime}\;.
\end{array}
\end{equation}
Each linear expression $f_{j}$  and $f_{j}'$
is assigned to $j$-th vertex of the left and right graphs
respectively.
Into each $f_{j}$ and $f_j'$ there involved four
linear variables from the sites surrounding the vertex.
Coefficients in $f_{j}$ and $f_j'$ are the vertex variables,
assigned with $j$-th vertex. Arrangement of the
vertex variables and the site variables with
respect to the arrows of the lines is the same
for any vertex (just look attentively at the linear relations
and the figure).
Each expression $f_{j}$ and $f_j'$ 
becomes the equation, $f_{j}=0$, $f_j'=0$,
and thus two sets of linear equations appear.

\begin{theorem}
Two systems,
(\ref{leftset}) and (\ref{rightset}) are
linearly equivalent (after excluding extra
linear variables $\varphi_h$ and $\varphi_a$) 
iff $\u_j',\w_j'$ are connected with $\u_j,\w_j$
via (\ref{R-map}).
\end{theorem}

\begin{figure}[tb]
\centering
\setlength{\unitlength}{0.25mm} 
\thicklines
\begin{picture}(450,210)
\put(125,0){
\begin{picture}(200,210)(0,-10)
\put( 150 ,   0 ){\vector( 0,1){200}}
\put(   0 , 150 ){\vector( 1,0){200}}
\put(   0 , 200 ){\vector(1,-1){200}}
\put( 150 , 150 ){\circle*{10}}
\put(  50 , 150 ){\circle*{10}}
\put( 150 ,  50 ){\circle*{10}}
\put(50,70){\scriptsize $\varphi_{3,a,b}$}
\put(-20,163){\scriptsize $\varphi_{2,a,b-1}$}
\put(155,-10){\scriptsize $\varphi_{2,a-1,b}$}
\put(107,110){\scriptsize $\varphi_{1,a,b}$}
\put(80,170){\scriptsize $\varphi_{3,a+1,b}$}
\put(170,90){\scriptsize $\varphi_{3,a,b+1}$}
\put(170,170){\scriptsize $\varphi_{2,a,b}$}
\end{picture}}
\end{picture}
\caption{Accessories of the lattice: cocurents.}
\label{fig-cocurrents}
\end{figure}
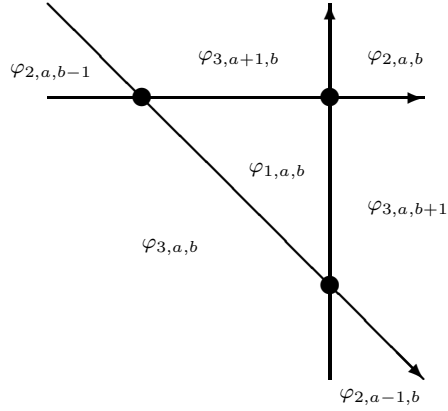

Turn now to the kagome lattice on the torus.
Linear forms $f_{j,a,b}$ and and corresponding
linear equations $f_{j,a,b}=0$ are to be introduced
for all vertices of the lattice.
\begin{equation}\label{linear}
\ds\begin{array}{@{}r@{}l}
\ds 0\;=\;f_{1,a,b}\;
\stackrel{\mbox{\scriptsize def}}{=}\;{}& \ds
\varphi_{3,a+1,b}\;-\;
\varphi_{2,a,b}\,\cdot\,\u_{1,a,b}\;+\;
\varphi_{1,a,b}\,\cdot\,\w_{1,a,b} \\[2pt]
&\ds {}+\;
\varphi_{3,a,b+1}\,\cdot\,
\kappa_1\,\u_{1,a,b}\,\w_{1,a,b}\;,\\
\\
\ds 0\;=\;f_{2,a,b}\;
\stackrel{\mbox{\scriptsize def}}{=}\;{}& \ds
\varphi_{1,a,b}\;-\;
\varphi_{3,a,b+1}\,\cdot\,\u_{2,a,b}\;+\;
\varphi_{3,a,b}\,\cdot\,\w_{2,a,b} \\[2pt]
&\ds {}+\;
\varphi_{2,a-1,b}\,\cdot\,
\kappa_2\,\u_{2,a,b}\,\w_{2,a,b}\;,\\
\\
\ds 0\;=\;f_{3,a,b}\;
\stackrel{\mbox{\scriptsize def}}{=}\;{}& \ds
\varphi_{3,a+1,b}\;-\;
\varphi_{1,a,b}\,\cdot\,\u_{3,a,b}\;+\;
\varphi_{2,a,b-1}\,\cdot\,\w_{3,a,b} \\[2pt]
&\ds {}+\;
\varphi_{3,a,b}\,\cdot\,
\kappa_3\,\u_{3,a,b}\,\w_{3,a,b}\;.
\end{array}\end{equation}
What these writings mean. 
The linear objects $\varphi^{}_{j,a,b}$, appeared in these
relations, are assigned to the sites of the kagome lattice
as it is shown in Figure~\ref{fig-cocurrents}. 
Each linear expression $f_{j,a,b}$ 
is assigned to $(j,a,b)$-th vertex of the kagome lattice, 
and into $f_{j,a,b}$ there involved four
linear variables from the sites surrounding the vertex.
Coefficients in $f_{j,a,b}$ are the vertex variables,
assigned with $(j,a,b)$-th vertex. Arrangement of the
vertex variables and the site variables with
respect to the arrows of the lines is the same
for any vertex.
Each expression $f_{j,a,b}$ becomes the equation, $f_{j,a,b}=0$,
and thus the set of linear equations appears.

The linearity of
$\varphi^{}_{j,a,b}$ allows one to impose the
quasiperiodical boundary conditions for them:
\begin{equation}\label{phi-AB}
\ds
\varphi_{\alpha,a-M,b}\;=\;
\varphi_{\alpha,a,b}\,\cdot\,A\;,\;\;\;
\varphi_{\alpha,a,b-M}\;=\;
\varphi_{\alpha,a,b}\,\cdot\,B\;.
\end{equation}
The coefficients of the linear system form a $3\,M^2\times
3\,M^2$ matrix ${\bf L}$, depending on the dynamical
variables and the quasimomenta $A,B$. The following set of
propositions was proved in \cite{s-3dsympl,s-qem}:
\begin{prop}
The determinant of ${\bf L}$ is a Laurent polynomial
of $A,B$,
\begin{equation}
\ds\det{\bf L}\;=\;\sum_{\alpha,\beta\in\Pi}\;\;
\widetilde{J}_{\alpha,\beta}\;A^\alpha\,\cdot\,B^\beta\;,
\end{equation}
where domain $\Pi$ is described below.
Being normalized in any way,
\begin{equation}
J(A,B)\;=\;{\det{\bf L}\over
\widetilde{J}_{\alpha_0,\beta_0}}\;,
\end{equation}
functional $J(A,B)$ is the generating functions for the
integrals of motion,
\begin{equation}
\ds
J(A,B)\;=\;\sum_{\alpha,\beta\in\Pi}\;\;
J_{\alpha,\beta}(\u,\w)\;
A^\alpha\,\cdot\,B^\beta\;,
\end{equation}
\begin{equation}
\ds
(U\,\circ\,J_{\alpha,\beta})(\u,\w)\;=\;
J_{\alpha,\beta}(\u,\w)\;.
\end{equation}
\end{prop}
\begin{prop}
Domain $\Pi$ in the decomposition of $J(A,B)$ 
is the following hexagon:
\begin{equation}\label{Pi}
\ds\Pi\;\;:\;\;
-M\;\leq\;\alpha\;\leq\;M\;,\;\;
-M\;\leq\;\beta\;\leq\;M\;,\;\;
-M\;\leq\;\alpha+\beta\;\leq\;M\;,
\end{equation}
where $M$ is the spatial size of the kagome lattice.
According to the Riemann-Hurwitz theorem, the genus of the
curve $\Gamma$ : $J(A,B)\;=\;0$ is
\begin{equation}
\ds g\;=\;3\,M^2\;-\;3\,M\;+\;1\;.
\end{equation}
Complete number of the integrals of motion is $3\,M^2+1$,
and one can choose exactly $3\,M^2$ involutive between them.
\end{prop}
Perimeter of the hexagon $\Pi$ is formed by $6M$ points,
\begin{equation}\label{perimeterpoints}
J_{M-n,n}\;,\;J_{-n,n-M}\;,\;
J_{M,-n}\;,\;J_{-M,M-n}\;,\;
J_{n-M,M}\;,\;J_{n,-M}\;,
\end{equation}
where $n=0,...,M$. These perimeter integrals 
are not independent. Let
\begin{equation}\label{XYZ}
\ds\begin{array}{@{}ccl}
\ds X_{\alpha} & = & \ds
\prod_{\sigma}\;\;\u^{-1}_{2,\alpha-\sigma,\sigma}\,
\u^{-1}_{3,\alpha-\sigma,\sigma}\;,\\
&&\\
\ds Y_{\beta} & = & \ds (-)^M\;
\prod_{\sigma}\;\;
\u{}_{1,\beta,\sigma}\,\w^{-1}_{3,\beta,\sigma}\;,\\
&&\\
\ds Z_{\gamma} & = & \ds
\prod_{\sigma}\;\;
\w^{}_{1,\sigma,\gamma}\,\w^{}_{2,\sigma,\gamma}\;.\\
\end{array}\end{equation}
Each of these expressions corresponds naturally to its line,
$X_\alpha$ to $x_\alpha$ etc., see Figure~\ref{fig-lines}. 
The perimeter integrals (\ref{perimeterpoints}) are some
the symmetrical polynomials of $X_\alpha$, or
$Y_\beta$, or $Z_\gamma$.
The Poisson brackets for $X_\alpha,Y_\beta,Z_\gamma$ 
$\forall \alpha,\beta,\gamma$ are
\begin{equation}
\ds
\{X_\alpha,Y_\beta\}\;=\;X_\alpha\,Y_\beta\;,\;\;
\{Y_\beta,Z_\gamma\}\;=\;Y_\beta\,Z_\gamma\;,\;\;
\{Z_\gamma,X_\alpha\}\;=\;Z_\gamma\,X_\alpha\;.
\end{equation}
It is useful to extract common convolutive parts
from $X_\alpha,Y_\beta,Z_\gamma$,
\begin{equation}
\ds
X_\alpha\;=\;X\,\cdot\,j(X_\alpha)\;,\;\;
Y_\beta\;=\;Y\,\cdot\,j(Y_\beta)\;,\;\;\;
Z_\gamma\;=\;Z\,\cdot\,j(Z_\gamma)\;,
\end{equation}
where
\begin{equation}
\ds \{X,Y\}\;=\;X\,Y\;,\;\;\;\mbox{etc.}
\end{equation}
Now we can rewrite $J(A,B)$ in its final invariant form:
\begin{equation}\label{spectral}
\ds J\;=\;
\sum_{\alpha,\beta,\gamma\,\in\,\Pi'}\;\;\;
j_{\alpha,\beta,\gamma}\;\;
(Z\,A)^\alpha\;(Y\,B^{-1})^{\beta}\;
\left(X\,{B\over A}\right)^\gamma\;,
\end{equation}
where $\Pi'$ is three squares of the cube:
\begin{equation}
\ds\Pi'\;\;:\;\;
0\;\leq\;\alpha,\beta,\gamma\;\leq M\;,\;\;\;
\mbox{at least one of}\;\;\alpha,\beta,\gamma\;\;
\mbox{iz zero}\;.
\end{equation}
Now we may describe the complete set of $3\,M^2$ involutive
integrals: $g=3M^2-3M+1$ functionals
$j_{\alpha,\beta,\gamma}$, corresponding to the inner points
of $\Pi'$, $3\,M-3$ independent projective $j(X_\alpha)$,
$j(Y_\beta)$ and $j(Z_\gamma)$, one $X\cdot Y\cdot Z$
(this is the center for $X,Y,Z$), and
any finally -- any other single function of $X$, $Y$, $Z$.

Running ahead, for what purpose else one needs the perimeter
integrals (\ref{perimeterpoints})? \linebreak[4]
$J(A,B)=0$ is an algebraic curve $\Gamma$, 
and we will be interested
in the divisors $(A)$ and $(B)$ of the algebraic
functions $A=A(P)$ and $B=B(P)$, $P\in\Gamma$.
In general the perimeter
integrals describes the divisors $(A)$ and $(B)$,
and after a bit cumbersome calculations we have obtained
the following description:
\begin{equation}\label{divA}
\ds\begin{array}{@{}ccl}
\ds (A)_0 &:&\ds\left\{\begin{array}{cl}
\ds A\;=\;0\;,& \ds {A\over B}\;=\;X_\alpha\;,\;\;\;\alpha\in Z_M\;,\\
&\\
\ds A\;=\;0\;,& \ds B\;=\;Y_\beta\;,\;\;\;\beta\in Z_M\;,
\end{array}\right.\;\\
&&\\
\ds (A)_\infty &:&\ds\left\{\begin{array}{cl}
\ds A\;=\;\infty\;,&\ds {A\over B}\;=\;
{X_\alpha\over(\kappa_2\kappa_3)^M}\;,\;\;\;\alpha\in Z_M\;,\\
&\\
\ds A\;=\;\infty\;,&\ds B\;=\;\left({\kappa_1\over\kappa_3}\right)^M
Y_\beta\;,\;\;\;\beta\in Z_M\;,
\end{array}\right.\;
\end{array}\end{equation}
and
\begin{equation}\label{divB}
\ds\begin{array}{@{}ccl}
\ds (B)_0 &:&\ds\left\{\begin{array}{cl}
\ds B\;=\;0\;,& \ds {A\over B}\;=\;X_\alpha\;,\;\;\;\alpha\in Z_M\;,\\
&\\
\ds B\;=\;0\;,& \ds A\;=\;{1\over (\kappa_1\kappa_2)^M Z_\gamma}
\;,\;\;\;\gamma\in Z_M\;,
\end{array}\right.\;\\
&&\\
\ds (B)_\infty &:&\ds\left\{\begin{array}{cl}
\ds B\;=\;\infty\;,&\ds {A\over B}\;=\;
{X_\alpha\over(\kappa_2\kappa_3)^M}\;,\;\;\;\alpha\in Z_M\;,\\
&\\
\ds B\;=\;\infty\;,&\ds A\;=\;{1\over Z_\gamma}
\;,\;\;\;\gamma\in Z_M\;,
\end{array}\right.\;
\end{array}\end{equation}
In these formulae it is supposed that the spectral
parameters $\kappa_j$ are the same for all lattice,
but the structure of $X_\alpha$, $Y_\beta$ and $Z_\gamma$,
associated with the lines $x_\alpha$, $y_\beta$ and
$z_\gamma$ makes obvious the situation when $\kappa_j$
depend on the lines numbers.

\section{The curve and solution}

Turn at last to the algebraic geometry. We deal with
the algebraic curve $\Gamma$, defined by
$J(A,B)\;=\;0$. $A$, $B$ and $C\;=\;A/B$ are the meromorphic
functions on $\Gamma$, their divisors
are already obtained in the previous section,
and (\ref{divA},\ref{divB}) may be rewritten in decent
notations as
\begin{equation}\label{AB}
\ds\begin{array}{@{}ccc}
\ds(A) & = & \ds\sum_{\alpha}\;\;
\left(P_{x_\alpha}^{+}\;-\;P_{x_\alpha}^{-}\right)\;+\;
\sum_{\beta}\;\;
\left(P_{y_\beta}^{+}\;-\;P_{y_\beta}^{-}\right)\;,\\
&&\\
\ds(B) & = & \ds\sum_{\alpha}\;\;
\left(P_{x_\alpha}^{+}\;-\;P_{x_\alpha}^{-}\right)\;-\;
\sum_\gamma\;\;
\left(P_{z_\gamma}^{+}\;-\;P_{z_\gamma}^{-}\right)\;,
\end{array}\end{equation}
where the notion of the points $P^\pm_{x_\alpha}$, 
$P^\pm_{y_\beta}$ and $P^\pm_{z_\gamma}$ comes
from (\ref{divA},\ref{divB}) transparently. 
Remarkable feature of these divisors is
the natural correspondence between $P^\pm$ and
the lines $x_\alpha$, $y_\beta$, $z_\gamma$.

Return now to linear system (\ref{linear}). Solving it, one
may put one of $\varphi_{j,a,b}$ to be unity, then all other
linear variables become some meromorphic functions on
$\Gamma$. Our next interest is a common pole divisor of all
$\varphi_{j,a,b}$. Denote this divisor as $\widetilde{\cal
D}$, its degree may be calculated:
\begin{equation}
\ds\mbox{deg}\;\widetilde{\cal D}\;=\;3\,M^2\;+\;1\;.
\end{equation}
This calculation is based directly on the dimension of linear
system and needs no comments.
The set $\varphi_{j,a,b}$ is the unique solution of the linear
system, so due to the Riemann-Roch theorem the dimension of
the linear space of the holomorphic functions with the pole
divisor $\widetilde{\cal D}$ is
\begin{equation}
\ds\mbox{dim}\;L(\widetilde{\cal D})\;=\;
\mbox{deg}\;\widetilde{\cal D}
\;-\;g\;+\;1\;=\;3\,M\;+\;1\;.
\end{equation}
This means that one may restore $\varphi_{j,a,b}$ as a
function on $\Gamma$ via just $3\,M$ points of positive
${\cal D}_{j,a,b}^\prime$,
\begin{equation}
\ds(\varphi_{j,a,b})\;+\;\widetilde{\cal D}\;=\;
{\cal D}^{(0)}_{j,a,b}
+{\cal D}_{j,a,b}^\prime\;,\;\;\;
\mbox{deg}\;{\cal D}^{(0)}_{j,a,b}\;=\;g\;,\;\;\;
\mbox{deg}\;{\cal D}_{j,a,b}^\prime\;=\;3\,M\;,
\end{equation}
where positive ${\cal D}_{j,a,b}^{(0)}$ may be restored
unambiguously. Obviously, the meaning of the quasimomenta $A$
and $B$
implies that we may choose the lines $x_{\alpha_0}$,
$y_{\beta_0}$ and $z_{\gamma_0}$, where
the quasimomenta appear, in any way. 
Therefore ${\cal D}_{\p}^\prime$,
$\p\equiv(j,a,b)$, are all governed by the same points as
form the divisors $(A)$ and $(B)$.

The same is valid for $\widetilde{\cal D}$:
\begin{equation}
\ds\widetilde{\cal D}\;=\;
\widetilde{\cal D}^{(0)}\;+\;\widetilde{\cal D}^\prime\;,
\end{equation}
where $\widetilde{\cal D}^\prime$, 
$\mbox{deg}\;\widetilde{\cal D}^\prime=3\,M$ 
is also governed by the points
of $(A)$ and $(B)$. Thus for any $\p=(j,a,b)$ the decomposition
arises:
\begin{equation}\label{phi-div}
\ds
(\varphi_{\p})\;=\;(\varphi_{\p})_0-(\varphi_{\p})_\infty\;=\;
{\cal D}_{\p}^{(0)}-\widetilde{\cal D}^{(0)}+
{\cal D}_{\p}^\prime-\widetilde{\cal D}^\prime\;,
\end{equation}
such that we can trace the following simple part of
$(\varphi_{\p})$:
\begin{equation}\label{phi-div-3m}
\ds{\cal D}_{\p}\;=\;{\cal D}_{\p}^\prime\;-\;
\widetilde{\cal D}^\prime\;.
\end{equation}

Before we give concrete expressions for $\varphi_{\p}$
recall a couple of notations of the algebraic geometry,
see for example \cite{Mumford}.
For a given curve $\Gamma$ with normalized holomorphic
one-forms $\o$ it is defined its jacobian $\mbox{Jac}\;\Gamma$
and the theta -- functions on it. We'll use the conventional
notations for $\Gamma\;\mapsto\;\mbox{Jac}\;\Gamma$: for
${\cal D}$, $\mbox{deg}\;{\cal D}\;=\;0$, let
\begin{equation}
\ds I({\cal D})\;=\;
\int_{\Sigma:{\cal D}=\partial\Sigma}\;\;\;\o\;.
\end{equation}
Also for $P,Q\;\in\;\Gamma$ denote their prime form as $E(P,Q)$,
\begin{equation}
\ds E(P,Q)\;\sim\;\Theta_\delta\left(\int_P^Q\;\;\o\right)\;,
\end{equation}
where the subscript $\delta$ means the nonsingular odd
theta characteristic of the curve (see \cite{Mumford}).
Let the same symbol $E$ stands for a product of
the prime forms 
\begin{equation}
\ds E(P,{\cal D})\;=\;\prod_{Q\in{\cal D}}\;\;E(P,Q)\;,\;\;\;
{\cal D}\;>\;0\;.
\end{equation}

\begin{theorem}
With the notations introduced, the expression for
$\varphi_{\p}$ as the meromorphic function on $\Gamma$ with
the divisor given by (\ref{phi-div},\ref{phi-div-3m}) is
\begin{equation}\label{phi-expression}
\ds
\varphi_{\p}\;=\;\varphi_{\p}(P)\;=\;
\varphi_{\p}^{(0)}\;\;
{\Theta({\bf z}+I(P-P_0+{\cal D}_{\p}))\over
\Theta({\bf z}+I(P-P_0))}\;\;
{E(P,({\cal D}_{\p})_0)\over E(P,({\cal D}_{\p})_\infty)}\;,
\end{equation}
where $P\in\Gamma$, vector ${\bf z}\in C^g$ and
$P_0\in \Gamma$ are some auxiliary parameters in the
parametrization (\ref{phi-expression}),
$\varphi_{\p}^{(0)}$ are constants (i.e. do not depend neither
on $P$ nor on ${\bf z}$),
$\Theta({\bf z}+I(P-P_0))$ corresponds to $g$ poles from
$\widetilde{\cal D}^{(0)}$, and so on. 
\end{theorem}
Actually eq. (\ref{phi-expression}) is the consequence
of the Riemann-Roch theorem, see \cite{korepanov-diss,Mumford}.

Further it is simpler to cancel $\Theta({\bf z}+I(P-P_0))$
from all $\varphi_{\p}$ and deal with the holomorphic with
respect to ${\bf z}$ functions.

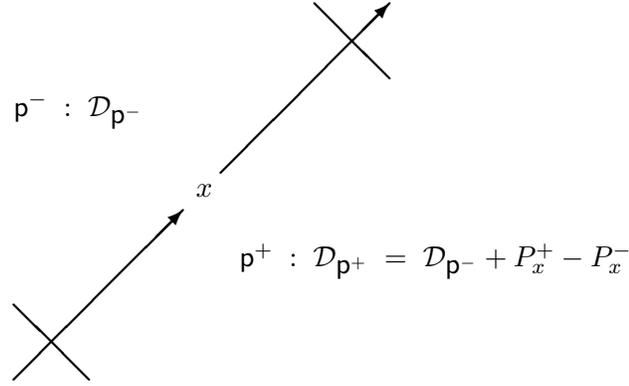
\begin{figure}[tb]
\centering
\setlength{\unitlength}{0.25mm}
\thicklines
\begin{picture}(350,200)
\put(75,0){
\begin{picture}(200,200)
\put(   0 ,   0 ){\vector( 1,1){90}}
\put( 110 , 110 ){\vector(1,1){90}}
\put(  40 ,   0 ){\line(-1,1){40}}
\put( 160 , 200 ){\line(1,-1){40}}
\put(  97 ,  97 ){$x$}
\put(   0 , 140 ){$\p^-\;:\;{\cal D}_{\p^-}$}
\put( 120 ,  60 ){$\p^+\;:\;{\cal D}_{\p^+}\;=\;{\cal D}_{\p^{-}}
+P_x^{+}-P_x^{-}$}
\end{picture}}
\end{picture}
\caption{Sites separated by a segment of line 
 $x$ between two vertices and theirs divisors.}
\label{fig-divisor-line}
\end{figure}

Describe now the explicit way of assigning the divisors
${\cal D}_{\p}$ (\ref{phi-div-3m})
to the sites $\p=(j,a,b)$ of the kagome lattice.

To each line $x_\alpha$, $y_\beta$ and $z_\gamma$ of the
lattice the pair $P^{+}_{..}-P^{-}_{..}$ is assigned, see
(\ref{AB}) and the remarks at the beginning of this section. 
Let a segment of oriented line $x$
separates two sites, $\p^{-}$ on the left and $\p^{+}$ on
the right according to the orientation of $x$, see 
Figure~\ref{fig-divisor-line}. According to our previous
considerations, the divisors ${\cal D}_{\p^+}$ and ${\cal
D}_{\p^-}$, assigned to $\p^+$ and $\p^-$ respectively, obey
\begin{equation}\label{div-game}
\ds
{\cal  D}_{\p^+}^{}\;-\;{\cal D}_{\p^-}^{}\;=\;
P_x^{+}\;-\;P_x^{-}\;,
\end{equation}
where $P_x^{+}$ and $P_x^{-}$ is the pair assigned to line $x$.
Starting from any site $\p_0$ on the lattice 
and using this procedure, we may 
define all ${\cal D}_{\p}$ up to the divisors 
of the algebraic functions $(A)$ and $(B)$.
Such system of the divisors was introduces by I. Korepanov in 
\cite{korepanov-diss}, i.e. the divisor rules we've obtained
(Figure~\ref{fig-divisor-line}, eq.~(\ref{div-game}))
coincide formally with that of the second type linear problem
(see the introduction).

\begin{figure}[tb]
\centering
\setlength{\unitlength}{0.25mm} 
\thicklines
\begin{picture}(450,240)
\put(125,0){
\begin{picture}(200,240)(0,-20)
\put(   0 ,   0 ){\vector( 1,1){200}}
\put( 200 ,   0 ){\vector(-1,1){200}}
\put( 100 , 100 ){\circle*{10}}
\put(  82 , 165 ){$b\,:\,{\cal D}_b$}
\put(  82 ,  25 ){$c\,:\,{\cal D}_c$}
\put(   0 ,  95 ){$a\,:\,{\cal D}_a$}
\put( 155 ,  95 ){$d\,:\,{\cal D}_b$}
\put( -20,-20){$x\,:\,{\cal D}_{x}^{}$}
\put( 190,-20){$y\,:\,{\cal D}_{y}^{}$}
\put(-20, 210){$y\,:\,{\cal D}_{y'}^{}$}
\put(190, 210){$x\,:\,{\cal D}_{x'}^{}$}
\end{picture}}
\end{picture}
\caption{Divisor notations around a vertex.}
\label{fig-divisor-vertex}
\end{figure}
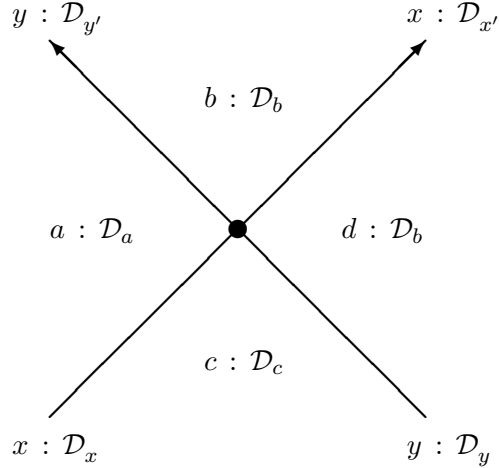

Introduce also a divisor of the edge of $x$ separating ${\p}^{+}$
and $\p^{-}$:
\begin{equation}
\ds
{\cal D}_{x}^{}\;=\;{\cal D}_{\p^-}^{}\;+\;P_{x}^{+}\;-\;P_0\;=\;
{\cal D}_{\p^+}^{}\;+\;P_{x}^{-}\;-\;P_0\;,
\end{equation}
where $P_0$ is the same point as in (\ref{phi-expression}).
The meaning of ${\cal D}_{x}^{}$ is following:
\begin{equation}\label{div-edge}
\ds
{\varphi_{\p^{-}}^{}\over\varphi_{\p^{+}}^{}}\;({\bf z}
\;=\;{\delta}\;-\;I({\cal D}_{x}^{}))\;=\;
{\varphi_{\p^{-}}^{(0)}\over\varphi_{\p^{+}}^{(0)}}\;.
\end{equation}

Consider further a vertex formed by two lines $x$ and $y$
surrounded by the sites $a,b,c,d$ as it is shown in 
Figure~\ref{fig-divisor-vertex}. The linear relation for it is
\begin{equation}\label{lin-gen}
\ds\varphi_{a}\;-\;\varphi_{b}\,\cdot\,\u\;+\;
\varphi_{c}\,\cdot\,\w\;+\;\varphi_{d}\,\cdot\,\kappa\,\u\,\w
\;=\;0\;.
\end{equation}
With the parametrization (\ref{phi-expression}) this
relation must be the identity both in $P$ and ${\bf z}$, so
that neither $\u$ nor $\w$ depend on $P$, but
$\u\;=\;\u({\bf z})$, $\w\;=\;\w({\bf z})$. This allows one
to find $\u({\bf z})$ and $\w({\bf z})$ immediately.

Eight divisors correspond to eight elements of the vertex.
For the site divisors we use obvious notations ${\cal
D}_{a}$, ${\cal D}_{b}$, ${\cal D}_{c}$ and ${\cal D}_{d}$.
Two lines involved, $x$ and $y$, are divided by the
intersection point into four edges with the divisors: ${\cal
D}_{x}$ and ${\cal D}_{x'}$ for $x$ line, and ${\cal D}_{y}$
and ${\cal D}_{y'}$ for $y$ line. According to divisor rules
(\ref{div-game}), the relations for all these eight divisors
may be written in the following form:
\begin{equation}
\ds\begin{array}{@{}ccccc}
\ds {\cal D}_{a }^{} & = & 
\ds {\cal D}_{x }^{}\;-\;P_{x}^{+}\;+\;P_0 & = & 
\ds {\cal D}_{y'}^{}\;-\;P_{y}^{+}\;+\;P_0\;,\\
&&&&\\
\ds {\cal D}_{b }^{} & = & 
\ds {\cal D}_{x'}^{}\;-\;P_{x}^{+}\;+\;P_0 & = & 
\ds {\cal D}_{y'}^{}\;-\;P_{y}^{-}\;+\;P_0\;,\\
&&&&\\
\ds {\cal D}_{c }^{} & = & 
\ds {\cal D}_{x }^{}\;-\;P_{x}^{-}\;+\;P_0 & = & 
\ds {\cal D}_{y }^{}\;-\;P_{y}^{+}\;+\;P_0\;,\\
&&&&\\
\ds {\cal D}_{d }^{} & = & 
\ds {\cal D}_{x'}^{}\;-\;P_{x}^{-}\;+\;P_0 & = & 
\ds {\cal D}_{y }^{}\;-\;P_{y}^{-}\;+\;P_0\;.
\end{array}\end{equation}
Testing (\ref{lin-gen}) for the quasiperiodicity on ${\bf
z}$ and taking zeros and poles of $\u({\bf z})$ and $\w({\bf
z})$ (relation (\ref{div-edge}) is useful for it), we get
unambiguously
\begin{equation}\label{solution}
\ds\u({\bf z})\;=\; \ds \u_0^{}\;\cdot\;
{\Theta\left({\bf z}\;+\;I({\cal D}_x)\right)\over
\Theta\left({\bf z}\;+\;I({\cal D}_{x'})\right)}\;,\;\;\;
\ds\w({\bf z})\;=\; \ds \w_0^{}\;\cdot\;
{\Theta\left({\bf z}\;+\;I({\cal D}_{y'})\right)\over
\Theta\left({\bf z}\;+\;I({\cal D}_{y})\right)}\;.
\end{equation}
Extra constant parameters obey
\begin{equation}
\ds\begin{array}{@{}ll}
\ds{\varphi_{a}^{(0)}\over\varphi_{c}^{(0)}}\;=\;
-\;\w_0^{}\,\cdot\,
{E(P_{x}^{+},P_{y}^{+})\over E(P_{x}^{-},P_{y}^{+})}\;,\;&
\ds{\varphi_{b}^{(0)}\over\varphi_{d}^{(0)}}\;=\;
\kappa\,\w_0^{}\,\cdot\,
{E(P_{x}^{+},P_{y}^{-})\over E(P_{x}^{-},P_{y}^{-})}\;,\\
&\\
\ds{\varphi_{a}^{(0)}\over\varphi_{b}^{(0)}}\;=\;
\u_0^{}\,\cdot\,
{E(P_{x}^{+},P_{y}^{+})\over E(P_{x}^{+},P_{y}^{-})}\;,\;&
\ds{\varphi_{c}^{(0)}\over\varphi_{d}^{(0)}}\;=\;
-\;\kappa\,\u_0^{}\,\cdot\,
{E(P_{x}^{-},P_{y}^{+})\over E(P_{x}^{-},P_{y}^{-})}\;.
\end{array}\end{equation}
These constants of the normalization, 
$\u_0$ and $\w_0$, are to be used to parametrize the
``gauge'' integrals of motion $X_\alpha$, $Y_\beta$, $Z_\gamma$,
and thus have no dynamical sense.

Actually (\ref{lin-gen}), as an identity on ${\bf z}$, is a
combination of two Fay's identities (see \cite{Mumford}). 
In the parametrization
of $\u$ and $\w$ the theta functions are assigned naturally
to the edges. These thetas are nothing but the triplet of
tau-functions and thus solve a system of trilinear
relations, see \cite{s-tau}.

With the parametrization in terms of the divisors, the
evolution becomes simple. It corresponds to the geometrical
shift of the lines generating the change of the divisors of
the sites. In general the lattice is heterogeneous and
non-equidistant and thus there the formula for
$\u,\w$ is rather geometrical, because of ``time'' means 
a concrete geometrical configuration. 
But $M$-step evolution $\mbox{\sf
S}\;=\;\U^M$ has the common form for all $\u,\w$:
\begin{equation}\label{period-T}
\ds
[\u,\w]\;=\;[\u({\bf z}),\w({\bf z})]\;
\stackrel{\mbox{\sf S}}{\longrightarrow}\;
[\u({\bf z}+{\bf T}),\w({\bf z}+{\bf T})]\;,
\end{equation}
where the vector of the periods
\begin{equation}\label{period-TT}
\ds
{\bf T}\;=\;I({\cal T})\;,
\end{equation}
is the same for all three ``time'' directions:
\begin{equation}
\ds
{\cal T}\;=\;\sum_\alpha\;P_{x_\alpha}^{+}-P_{x_\alpha}^{-}\;=\;
\sum_\alpha\;P_{y_\alpha}^{-}-P_{y_\alpha}^{+}\;=\;
\sum_\alpha\;P_{z_\alpha}^{+}-P_{z_\alpha}^{-}\;,
\end{equation}
where, surely, 
the equality signs mean the equivalence of the divisors.
 
\section{Conclusion}

In this conclusion we would like to discuss possible applications
of all the considerations made in this paper. 
Usually the algebraic geometry gives rather formal results,
which are hard to apply to calculate something ``physical''.
Nevertheless one may use the considerations above
in order to make some important conclusions for the
quantum evolution model \cite{s-qem}.

The quantum evolution model is based on the quantum analogue
of the map $\R$ (\ref{R-map}) for the local Weyl algebra
replacing the local Poisson algebra (\ref{localPoisson}),
\begin{equation}\label{Weylalgebra}
\ds \u_j^{}\*\w_j^{}\;=\;q\;\;\w_j^{}\*\u_j^{}\;,
\end{equation}
where $q$ is the commonly accepted parameter of the 
quantum deformation. 
The form of the map $\R$ and the linear system 
in the quantum case differ very slightly form
(\ref{R-map}) and (\ref{leftset},\ref{rightset}),
see \cite{s-qem} for the details. The determinant
in the quantum case is well defined and also gives
the complete set of the quantum integrals of motion.
With a great effort we do not fall into speculations 
concerning a quantum jacobian.

Instead, turn to the case when $q$ is $N$-th root of unity.
For these $q$ the Weyl algebra (\ref{Weylalgebra}) has
the finite dimensional representations, 
so that $N$-th powers $\u_j^N$ and $\w_j^N$ are centres,
i.e. the parameters of the finite dimensional representation.
The map for $\u_j^N$ and $\w_j^N$ coincides
with the functional map (\ref{R-map}) up to
$\kappa_j\mapsto\kappa_j^N$, i.e. exactly the case considered.
Actually the quantum map factorizes into
a finite dimensional part (which is the vertex $R$ matrix
for a Zamolodchikov -- Bazhanov -- Baxter -- type model, giving
the Boltzmann weights 
for a statistical mechanics modelling) {\em times} the functional
part for $N$-th powers. The evolution operator
factorizes also in this way.
The quantum $S$ matrix is a high power of the 
quantum one-step evolution operator, 
and so in order to factorize it one has
to carry out all the functional parts outside.
Doing this, one changes the parameters of the finite dimensional
one-step evolution operators. This situation corresponds
to the consistent changes of the Boltzmann
weights parameters from one time layer to the other.
In two dimensions the simplest such models are known 
as the checkerboard models, and the simplest three dimensional
chess model was described in \cite{mss-elliptic,bmss-elliptic}.
Note, the parameters $\u^M,\w^N,\kappa$ live on the lattice,
i. e. $\u_V^N$, $\w_V^N$, $\kappa_V$ are 
assigned to the vertex $V$ as well as the operators $\u_V,\w_V$.
Actually, for given spatial size of the kagome lattice $M$,
one may consider another effective spatial size $M'$ of the
lattice of parameters, such that $M$ is divisible by $M'$.
Small $M'$ correspond to the small heterogeneity of the
lattice, and $M'=1$ corresponds to the spatial homogeneity
of the parameters.

In general the results of this paper
would help one to parametrize the consistent evolution of 
the parameters of the finite dimensional one-step
evolution operators or the transfer matrices.

\subsection*{Acknowledgement} 
I would like to thank I. Korepanov,
R. Kashaev, G. Pronko and Yu. Stroganov for many fruitful
discussions. The work was supported by the RFBR grant No. 98-01-00070.

\label{sergeev_lp}

\end{document}